\documentclass[preprint,preprintnumbers,amsmath,amssymb,,nofootinbib,12pt]{revtex4-2}
\usepackage{amsfonts}
\usepackage{amssymb}
\usepackage{latexsym}
\usepackage{graphicx}
\usepackage{verbatim}
\usepackage{rotating}
\usepackage{multirow}
\usepackage{booktabs}
\usepackage{blindtext}
\usepackage[colorlinks,allcolors=blue]{hyperref}
\usepackage[usenames,dvipsnames]{color}
\usepackage[active]{srcltx}
\newcommand{\re}[1]{(\ref{#1})}

\newcommand{\al}{\alpha}

\newcommand{\veps}{\varepsilon}
\newcommand{\Si}{\Sigma}

\newcommand{\rar}{\rightarrow}

\newcommand{\non}{\nonumber}

\begin{document}
\title{Towards the ``puzzle" of the Chromium dimer Cr$_2$: predicting the Born-Oppenheimer rovibrational spectrum}
\author{Horacio~Olivares-Pil\'on}
\email{horop@xanum.uam.mx}
\affiliation{Departamento de F\'isica, Universidad Aut\'onoma Metropolitana-Iztapalapa,
Apartado Postal 55-534, 09340 M\'exico, D.F., Mexico}
\author{Daniel~Aguilar-D\'iaz}
\email{cbi2221801227@xanum.uam.mx}
\affiliation{Departamento de F\'isica, Universidad Aut\'onoma Metropolitana-Iztapalapa,
Apartado Postal 55-534, 09340 M\'exico, D.F., Mexico}
\author{Alexander~V.~Turbiner}
\email{turbiner@nucleares.unam.mx}
\affiliation{Instituto de Ciencias Nucleares, Universidad Nacional
Aut\'onoma de M\'exico, Apartado Postal 70-543, 04510 M\'exico, D.F., Mexico}

\begin{abstract}

The experimentally-observed non-trivial electronic structure of the Cr$_2$ dimer
has made the calculation of its potential energy curve a theoretical challenge
in the last decades. By matching the perturbation theory at small internuclear
distances $R$ and the multipole expansion at large distances $R$ (supposedly both
of asymptotic nature), and by adding a few Rydberg-Klein-Rees (RKR) turning points,
extracted from experimental data by Casey-Leopold (1993),
the analytic form of the potential energy curve for the ground state $X^1\Si^+$
of the Cr$_2$ dimer is found {   for the first time} for the whole range of internuclear
distances $R$.
This has the form of a two-point Pad\'e approximant and provides an accuracy of 3-4 decimal
digits in 29 experimental vibrational energies. The resulting ground state $X^1\Si^+$
potential curve supports 19694
%19908
rovibrational states with a maximal vibrational number $\nu_\text{max}=104$ at zero angular momentum and with a maximal angular momentum $L_\text{max}=312$ with energies
$> 10^{-4}$ {\it hartree}, and additionally 218 weakly-bound states
(close to the dissociation limit) with energies $< 10^{-4}$ {\it hartree}.\\

%ArXiv: 2401.03259

\end{abstract}
\maketitle
%--------------------------------------------------------------------------
\section{INTRODUCTION}

%Intrigued by a  red mineral that had been discovered in a Siberian gold mine in 1766
%(called crocoite PbCrO$_4$ - lead chromate), Louis Nicholas Vauquelin (1763-1829) discovered
%and isolated the Chromium in 1798~\cite{JE:2001}.
%The history of chromium dimer begins with the discovery and understanding of
%metal-metal multiple bonds. In 1844 Eugène Melchior Peligot (1811-1890)
%synthesized Chromium(II) acetate Cr$_2$(OAc)$_4$(H$_2$O)$_2$ without
%knowing that he was the first to prepare the first compound with a metal-metal multiple
%bond, and it was not until 1970 that the structure  of this compound was solved
%crystallographically~\cite{EP:1844,JB:2017}.

The chromium dimer Cr$_2$ is made of 48 electrons and two Chromium nuclei, typically ${}^{52}Cr$. For decades it has been a challenge to study this dimer both
experimentally and theoretically. Even in its ground state, the Cr atom
has 6 out of its 24 electrons unpaired in the 3d and 4s orbitals (3d)$^5$(4s),
which signals a complex electronic structure. This implies that for the dimer
Cr$_2$, the metal bond involves twelve electrons forming six bonds, which are
weakly-bound, since the binding energy of Cr$_2$ is only $\sim 0.05$~{\it hartree}
at the equilibrium distance $R_{eq}\sim 3.17$ a.u.
As a result, chromium dimers are typically synthesized and studied under inert
conditions~\cite{JB:2017,FC:1970,DM:1982}.
%$\sim 1.44$~eV

In the 1970s, theoretical studies on the nature of multiple bonds in metals
were carried out, and the chromium dimer emerged as a key problem, due to
its complex electronic structure possibly leading to an unusual potential energy curve.
The formal bond of order 6 in Cr$_2$ has required a high level of theoretical accuracy
to describe the dynamic and non-dynamic correlations of the electrons. In 2022 Larsson et al.~\cite{HL:2022} compiled the main theoretical works giving an overview of
the different theoretical approaches developed over the last 40 years by numerous authors.
From this compilation it became explicitly clear that there is a striking, sometimes,
even qualitative difference in the potential energy curves
obtained using different theoretical approaches, mostly based on
{\it ab initio} calculations, see Fig.1 in \cite{HL:2022}.

A {   \it single} experimental work, where the vibrational energy transitions
in the ground electronic state of Cr$_2$ were studied, was due to
Casey-Leopold~\cite{CL:1993} (see also \cite{KA:1995}).
It measured 29 transition energies in three distinct regions (subbands) of the vibrational
spectra. Then by using the RKR procedure it found 29 pairs
of turning points and eventually constructed a RKR potential energy curve of these three subdomains of internuclear distances with $R \in$ (2.83, 6.14)~a.u.
\footnote{   It must be emphasized that in any {\it ab initio} calculation a number of discrete points on the potential curve are found only, which then are {\it interpolated} to obtain a continuous potential curve. In order to use it as the potential in the nuclear Schr\"odinger equation this curve must be {\it extrapolated} to the domains of large and small internuclear distances. Both procedures of interpolation and extrapolations lead to some uncertainties in the energy spectrum.} The work demonstrated
a quite unusual, exotic behavior of the obtained RKR potential energy curve in this domain. Recently, a reassignment of the vibrational quantum numbers in the energy subbands \cite{HL:2022}
extracted from the experimental data by Casey-Leopold~\cite{CL:1993} was suggested
\footnote{in this work we will follow this reassignment, which is in agreement with our calculations of vibrational spectrum}. {
This reassignment makes the RKR potential energy curve be in a certain quantitative
agreement with the theoretical calculations carried out in \cite{HL:2022} in the domain
$R \in$ (2.83, 6.14)~a.u.}\footnote{{Three highly advanced theoretical methods
were used in \cite{HL:2022}, see also Ref.26 therein, to calculate a potential energy curve
at some points in this finite domain. It should be noted that in the Born-Oppenheimer formalism
the knowledge of the potential (curve) in a finite domain does not allow to calculate
the rovibrational spectra in full generality reliably due to the impossibility of imposing
proper boundary conditions in the nuclear Schr\"odinger equation.}}.
However, there is still a significant disagreement with the
works of different authors about the spectroscopical constants and, in particular,
about the equilibrium distance $R_{eq}$ and dissociation energy $E_d$, see Fig.3
in \cite{HL:2022}.

\pagebreak

In this work, using the methodology recently developed thoroughly
in~\cite{AT:2022,PT:2023} and successfully applied to HeH, H$_2$ \cite{OT:2018},
LiH, He$_2^+$ \cite{AT:2022}, ClF \cite{OT:2022}, H(D,T)F \cite{AG-OP:2022},
H(D,T)Cl \cite{PT:2023},
which is based
on matching the perturbation theory at small distances and the
multipole expansion at large distances combined with the available experimental
and/or reliable {\it ab initio} calculation information {   (for a few electron cases)} at intermediate internuclear distances, a Born-Oppenheimer ground state potential energy
curve for Cr$_2$ is presented {   in an analytic form}. This curve should describe
the entire domain $R\in[0,\infty)$
with 3-5 {   significant digits (s.d.)} of accuracy. It is worth mentioning that the
exponentially small terms, which appear in the theory presented in~\cite{AT:2022} at large internuclear distances, are not considered in the present study, since we are interested only
in the ground state of the chromium dimer, while they become essential only for
the excited states.
%Hence, the obtained curve can be considered as phenomenological.
It must be emphasized that following our previous analysis
performed for Helium-like and Lithium-like atomic sequences as well as for the
H$_2^+$, H$_2$ and $H_3^+$ molecules, we assume the existence of a correction-free
domain for the rovibrational energies: the mass corrections $m_e/M_{nuclei}$,
relativistic corrections $v/c$ and QED corrections $\al^2$ contribute to
the 4th significant digit (and subsequent ones) in the energies. By
taking the obtained potential curve (the electronic term) as the potential
in the nuclear two-body Schr\"odinger equation the entire rovibrational spectrum
of the Cr$_2$ dimer is calculated for the first time.

Atomic units are used throughout the paper. Energies are assumed in {\it hartree}.

%----------------------------------------------------------------------------------%----------------------------------------------------------------------------------
%%%%%%%%%%%%%%%%%%%%%%%%%%%%%%%%%

\section{The ground state $X^1\Sigma^+$ of the Cr$_2$ molecule}

The Chromium dimer Cr$_2$ is composed of two Chromium atoms Cr ($Z_{\text{Cr}} = 24$).
The {   binding} energy $\tilde{E}$ is related to the total energy $E(R)$ by
\begin{equation}
    \tilde{E}(R)\ =\ E(R)-2E_{\text{Cr}} \ ,
\end{equation}
where $E_{\text{Cr}}=-1050.4$ {\it hartree} \cite{NIST:2022} is the ground state energy
of the Chromium atom Cr. The united atom limit of the Cr$_2$ dimer corresponds
to the Cadmium atom Cd, for which the ground state energy is taken equal to
$E_{\text{Cd}}=-5588.0$~{\it hartree}~\cite{NIST:2022}.
{   The binding energy at the equilibrium distance is equal to the dissociation
energy, $\tilde{E}(R_{eq})=E_d$}.

Following Bingel \cite{AB:1959}, for small internuclear distances
$R \rar 0$ the dissociation energy behaves as
\begin{equation}
    \tilde{E}\ =\ \frac{576}{R}\ +\ \veps_0\ +\ O(R^2)\ ,
\label{small_distances}
\end{equation}
where the first term is the Coulomb repulsion term of the nuclei $Z_{\text{Cr}}^2/R$ and
$\veps_0$ is the difference of the ground state energies of
the united atom and of the two separated Cr atoms:
$\veps_0=E_{\rm Ca} -2 E_{\rm Cr}\,=\,-3487.2$ {\it hartree}.
The behavior of the dissociation energy at large internuclear distances $R \rar \infty$
is given by~\cite{IGK:2006}
\begin{equation}
    \tilde{E}\ =\ -\frac{C_6}{R^6}\ +\ O\left(\frac{1}{R^8}\right)\ ,
\label{large_distances}
\end{equation}
where the van der Waals constant is $C_6=745.0 \pm 55 \ \text{a.u.}$~\cite{ZP:2004}.

In general, the analytic expression, which interpolates the domains of small
$R$~\re{small_distances} and large $R$~\re{large_distances},
is taken to be a ratio of two polynomials $P_N(R), Q_{N+5}(R)$ with difference in degrees equal
to 5 multiplied by $1/R$. This ratio is treated as a two-point Pad\'e approximant~\cite{AT:2022},
\[
    \mbox{Pade}[N/N+5](R)\ =\ \frac{P_N(R)}{Q_{N+5}(R)} \ ,
\]
where $N$ is the order of the approximant, which is chosen accordingly, see below.

The potential energy curve $E(R)_{\{3,2\}}$ (which coincides with the dissociation energy
$\tilde{E}(R)$), is presented for $N=6$ as
\begin{equation}
      E(R)_{\{3,2\}}\ =\ \frac{1}{R}\,
      \frac{576+\sum_{k=1}^{4}\,a_k\,R^k+a_5\,R^5-a_6\,R^6}
      {(1+\alpha_1\,R+\alpha_2\,R^2+\sum_{k=3}^{9}\,b_k\,R^k+b_{10}\,R^{10}+ b_{11}\,R^{11})}\ .
\label{E-Pade}
\end{equation}
Four constraints are imposed in order to reproduce the asymptotic limits~\re{small_distances}
and \re{large_distances}
\begin{eqnarray}
\label{cons1}
     \al_1 &=& \frac{a_1-\veps_0}{576}\ , \\
      \al_2 &=& \frac{-a_1\varepsilon_0+\varepsilon_0^2 + 576 a_2}{576^2}\ ,\non \\
       a_5 &=& -b_{10}\,C_6\ , \non\\
        a_6 &=&  b_{11}\,C_6\ , \non
\end{eqnarray}
{   where we choose $C_6=745.0 \ \text{a.u.}$, see below for discussion.}

\begin{figure}[ht]
\includegraphics[scale=1.70]{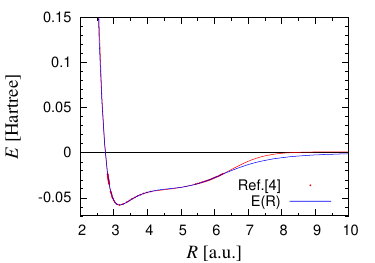}
\centering
\llap{\shortstack{%
        \includegraphics[scale=0.65]{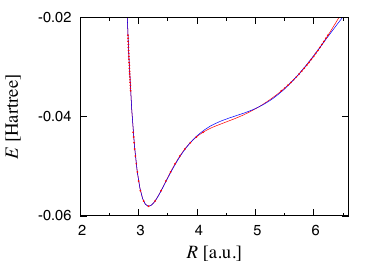}\\
        \rule{0ex}{1.4in}}
  \rule{0.4in}{0ex}}
\caption{Potential energy curve for the ground state $X^1\Si^+$
  of the dimer Cr$_2$ {\it  vs.} internuclear distance $R$:
  (i) the Pad\'e approximant~\re{E-Pade} (continuous blue line),
  (ii) experimental data \cite{HL:2022} (red bullets). The minimum
  $E_{min}=-0.058063$~{\it  {\it hartree}} is located at $R_{eq}= 3.17247 $~a.u.,
  both in agreement with~\cite{HL:2022} and~\cite{CL:1993}.
  The inset shows the vicinity of the minimum of the potential energy curve,
  where the experimental data situated, 
  (iii) potential curve from \cite{HL:2022} (continuous blue line).}
\label{fig:potCurve}
\end{figure}

The remaining twelve parameters in ~\re{E-Pade} are free. These parameters
are fixed by fitting some RKR experimental data~\cite{HL:2022}
in the domain $R\in [2.8, 6.1]$~a.u. After {   state-of-the-art, mostly manual} fitting, these parameters take the values
\begin{equation}
\begin{array}{rcr  rcr  rcr  rcr}
  a_1\ &=&\  34.95358\ , & a_2 &=& -5.681092\ ,  & a_3  &=& -21.92814\ ,\\
  a_4\ &=&\ -44.02492\ , & a_5 &=& 25.45038\ ,   & a_6  &=&  3.834068\ ,\\
  b_3\ &=&\  1.514311\ , & b_4 &=&  6.085325\ ,  & b_5  &=&  5.66723\ , \\
  b_6\ &=&\ -8.952521\ , & b_7 &=& -0.1508989\ , & b_8  &=&  1.492459\ , \\
  b_9\ &=&\ \ \ -0.16354\ ,\ &\ \ b_{11} &=& 0.0051464\ , &  &\
\label{eq:a_i,b_i}
\end{array}
\end{equation}
{   where all numbers are in  a.u.}
One can check that by inserting the parameters~\re{eq:a_i,b_i} and by imposing the
constraints~\re{cons1} in~\re{E-Pade}, the Taylor expansions at $R=0$ (in powers of $R$)
and $R=\infty$ (in powers of $1/R$) of $E(R)_{\{3,2\}}$~\re{E-Pade}, reproduce exactly
the first 3 terms in~\re{small_distances} and the first 2 terms
in~\re{large_distances}. Let us also mention that by making
comparison with~\cite{HL:2022} and~\cite{CL:1993}, one can see that
the position and the depth of the minimum of the potential $\tilde{E}(R)$ are
reproduced: $R_{eq}= 3.17247 $~a.u.,  $E_{min}=-0.058063$~{\it  hartree}
(see Figure~\ref{fig:potCurve}). In Table~\ref{tab:1} the comparison of the energy
values $\tilde{E}(R)$, resulting from~\re{E-Pade} at the turning points, see~\cite{CL:1993,HL:2022},
and experimental data for the ground state $X^1\Sigma^+$ of Cr$_2$
(see also Figure~\ref{fig:potCurve}) is presented.
The agreement is within 3-4 decimal digits.

{   It is worth noting that the dependence on the van der Waals constant $C_6$ of the potential curve is almost negligible (in reasonable limits) being inside of the error bars of the RKR-obtained turning points which is about $10^{-4}$ in relative units. Varying $C_6$ inside of the error bars $\pm 55$\,a.u., which is $\sim 7\%$, and refitting the parameters in the Pad\'e approximant (\ref{E-Pade}) leads to variations of the potential curve within the uncertainty of the experimental data \cite{CL:1993}: the potential curve changes in the 4th decimal digits in 1-2 portions in the physically important domain $R \in [2-8]$\,a.u.}

{   In Fig.~\ref{fig:potCurve} our potential energy curve (\ref{E-Pade}) with parameters (\ref{cons1}) and (\ref{eq:a_i,b_i}) is compared with the so-called extended Morse Oscillator (EMO) curve presented in \cite{HL:2022}. Both curves coincide in the domains which correspond to the experimentally observed bound states but differ significantly at $R \lesssim 2.5$\,a.u.,  $R \in [4 - 5.5]$\,a.u. and $R \gtrsim 6.3$\,a.u. This is understandable since}
\pagebreak

\noindent
{   the EMO potential curve predicts non-physical behavior at small and large internuclear
distances~\footnote{{   It decays exponentially at large distances.}}.
}
%-------------------------------------------------------------------------
%-------------------------------------------------------------------------

\subsection{Rovibrational spectra}

In the Born-Oppenheimer approximation, the rovibrational spectra
$E_{(\nu,L)}$ can be obtained by solving the nuclear radial Schr\"odinger equation
\begin{equation}
   \left(-\frac{1}{2\mu}\frac{d^2}{dR^2}+\frac{L(L+1)}{2\mu R^2}+V(R)\right)\psi(R)\ =\  E_{(\nu,L)}\, \psi(R)\ ,
\label{Hamiltonian}
\end{equation}
where $\nu$ and $L$ are the vibrational and rotational quantum numbers and
$\mu$ the reduced mass of the chromium dimer Cr$_2$. Since
the most abundant isotope of chromium is $^{52}$Cr,
the nuclear mass $m_{^{52}\text{Cr}}=94657.75308 $ \cite{GA:2003} is chosen.
Eq.~\re{Hamiltonian} is solved using the Lagrange-mesh method (LMM)
where the Schr\"odinger equation (\ref{Hamiltonian}) is discretized to a non-uniform lattice
with mesh points defined by the zeros  of the Laguerre polynomials~\cite{DB:2015}.
In order to calculate the vibrational energies at fixed $L$, a large
number of mesh points were considered (up to 1100). Calculations were carried out
at ICN-UNAM cluster Karen.

\begin{figure}[h]
	\centering
	\includegraphics[scale=2.4]{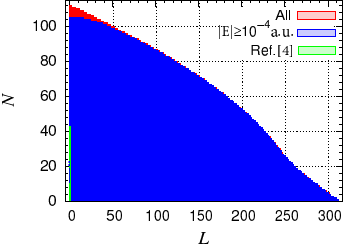}
\caption{Number of vibrational states for given algular momentum
    for the ground state X$^1\Sigma^+$
    of the Cr$_2$ dimer as a function of the angular momentum $L$. In blue the
    rovibrational states are indicated which are found with absolute accuracy
    $\sim 10^{-4}$ {\it  hartree}, in total there are 19694 states. In {\color{red} \bf red} the 218
    weakly-bound states with energies  $< 10^{-4}$ {\it  hartree} shown. In {\color{green} \bf green}
    the experimental data from~\cite{CL:1993,HL:2022} shown (highlighted for better visualization).}
\label{fig:rovibrational_states}
\end{figure}

As a result of solving equation~\re{Hamiltonian}, it is found that the potential energy curve
{   (\ref{E-Pade}) with parameters (\ref{cons1}) and (\ref{eq:a_i,b_i})}
for the ground state $X^1\Sigma^+$ of Cr$_2$ supports 112 vibrational states $E_{(\nu,0)}$ ($L=0$)
with $\nu=0,\dots,111$ ($\nu_{max}=111$). Among these vibrational levels, the states
with $\nu =\{105,106,\dots,111\}$ are weakly-bound, they are near the dissociation
threshold with energies smaller than $10^{-4}$ {\it  {\it  hartree}}, thus, beyond the accuracy
of the Born-Oppenheimer approximation. A comparison with the experimental results
is done for the vibrational states with $\nu =\{0,\dots,9,23,\dots,42\}$ presented
in~\cite{HL:2022}, see Table~\ref{tab:2}. These are in complete agreement. Total number of
vibrational energy states is sufficiently large to allow us to construct reliably
the partition function and thus to study the thermodynamical characteristics of
the non-rotating Cr$_2$ dimer, in particular.
In Figure~\ref{fig:rovibrational_states} the histogram of the rovibrational spectra
is presented. In total, there are 19912 rovibrational states with $\nu_{\text{max}} = 111$
and $L_{\text{max}} = 312$. Out of these there are 19694 (strongly)-bound rovibrational
states with energies $\gtrsim 10^{-4}$ {\it  {\it  hartree}}. {   It can be shown that
this number 19694  of rovibrational states remains the same when the van der Waals constant
$C_6$ is changed within its error bars \cite{ZP:2004}: $C_6=745 \pm 55$\,a.u.}

\begin{table}[h]
\caption{Potential energy curve $E(R)$~\re{E-Pade} of the ground
    state $X^1\Sigma^+$ of the Cr$_2$ dimer evaluated at the
    RKR style turning points~\cite{CL:1993,HL:2022}. For comparison,
    the results from~\cite{HL:2022} are shown in column 5.
}
\scalebox{1}{%
\begin{tabular}{cc|cc|c}
\hline\hline
\,\,\,$R_{\text{min}}$\hspace{0.5cm} & $R_{\text{max}}$ & $E(R_{\text{min}})$ & $E(R_{\text{max}})$ & Ref.~\cite{HL:2022}\\
\hline
3.0850& 3.2830& -0.0570& -0.0570& -0.0570 \\
%3.0320& 3.3847& -0.0549& -0.0550& -0.0549 \\
%3.0007& 3.4678& -0.0530& -0.0531& -0.0530 \\
2.9776& 3.5482& -0.0512& -0.0513& -0.0512 \\
%2.9601& 3.6260& -0.0495& -0.0497& -0.0495 \\
%2.9458& 3.7069& -0.0479& -0.0481& -0.0479 \\
2.9339& 3.7940& -0.0465& -0.0466& -0.0465 \\
%2.9241& 3.8872& -0.0452& -0.0453& -0.0452 \\
%2.9160& 3.9917& -0.0441& -0.0440& -0.0441 \\
2.9096& 4.1058& -0.0431& -0.0429& -0.0431 \\
%2.8630& 5.4517& -0.0347& -0.0349& -0.0348 \\
%2.8605& 5.5062& -0.0342& -0.0344& -0.0342 \\
2.8581& 5.5561& -0.0337& -0.0338& -0.0337 \\
%2.8555& 5.6078& -0.0331& -0.0332& -0.0331 \\
%2.8530& 5.6553& -0.0325& -0.0326& -0.0325 \\
2.8504& 5.7044& -0.0320& -0.0320& -0.0320 \\
%2.8480& 5.7497& -0.0314& -0.0314& -0.0314 \\
%2.8455& 5.7949& -0.0308& -0.0307& -0.0308 \\
2.8431& 5.8367& -0.0302& -0.0301& -0.0302 \\
%2.8407& 5.8786& -0.0296& -0.0295& -0.0296 \\
%2.8382& 5.9206& -0.0290& -0.0289& -0.0290 \\
2.8358& 5.9597& -0.0285& -0.0283& -0.0285 \\
%2.8334& 5.9990& -0.0278& -0.0277& -0.0279 \\
%2.8311& 6.0358& -0.0273& -0.0271& -0.0273 \\
2.8289& 6.0717& -0.0267& -0.0266& -0.0267 \\
%2.8267& 6.1067& -0.0261& -0.0260& -0.0260 \\
%2.8245& 6.1410& -0.0255& -0.0255& -0.0254 \\
2.8221& 6.1771& -0.0249& -0.0249& -0.0248 \\
%2.8200& 6.2100& -0.0243& -0.0244& -0.0242 \\
2.8179& 6.2424& -0.0237& -0.0239& -0.0236 \\
\hline\hline
\end{tabular}
}
\label{tab:1}
\end{table}
%%%%%%%%%%%%%%%%%%%%%%%
\begin{table}[h]
\caption{Vibrational energies $E_{(\nu, 0)}$ of the ground state
     $X^1\Sigma^+$ of the diatomic molecule Cr$_2$. For comparison,
     the results from~\cite{HL:2022} are shown in columns 2 and 5.
     The levels with $\nu=19$ and $21$ presented in ~\cite{HL:2022,CL:1993}
     are considered uncertain. States with assignment $\nu = {105, . . . , 111}$
     (marked by bold) are weakly-bound, they are close to the dissociation limit.}
\scalebox{0.85}{
\begin{tabular}{llc | llc | lc}
\hline\hline
$\nu$& Ref.~\cite{HL:2022}& $E_{\nu,0}$& $\nu$& Ref.~\cite{HL:2022}& $E_{\nu,0}$& $\nu$& $E_{\nu,0}$\\
\hline
0 &  -0.0570& -0.0570& 30& -0.0308& -0.0307&  60& -0.0131\\
1 &  -0.0549& -0.0549& 31& -0.0302& -0.0301&  62& -0.0120\\
2 &  -0.0530& -0.0530& 32& -0.0296& -0.0295&  64& -0.0110\\
3 &  -0.0512& -0.0512& 33& -0.0290& -0.0290&  66& -0.0100\\
4 &  -0.0495& -0.0496& 34& -0.0285& -0.0284&  68& -0.0090\\
5 &  -0.0479& -0.0480& 35& -0.0279& -0.0278&  70& -0.0081\\
6 &  -0.0465& -0.0466& 36& -0.0273& -0.0272&  72& -0.0072\\
7 &  -0.0452& -0.0453& 37& -0.0267& -0.0266&  74& -0.0064\\
8 &  -0.0441& -0.0441& 38& -0.0260& -0.0260&  76& -0.0056\\
9 &  -0.0431& -0.0431& 39& -0.0254& -0.0254&  78& -0.0048\\
10&         & -0.0421& 40& -0.0248& -0.0247&  80& -0.0042\\
11&         & -0.0413& 41& -0.0242& -0.0241&  82& -0.0035\\
12&         & -0.0406& 42& -0.0236& -0.0235&  84& -0.0029\\
13&         & -0.0400& 43&        & -0.0229&  86& -0.0024\\
14&         & -0.0395& 44&        & -0.0223&  88& -0.0019\\
15&         & -0.0389& 45&        & -0.0217&  90& -0.0015\\
16&         & -0.0384& 46&        & -0.0211&  92& -0.0011\\
17&         & -0.0379& 47&        & -0.0205&  94& -0.0008\\
18&         & -0.0374& 48&        & -0.0199&  96& -0.0006\\
19&  -0.0374& -0.0369& 49&        & -0.0194&  98& -0.0004\\
20&         & -0.0364& 50&        & -0.0188& 100& -0.0003\\
21&  -0.0362& -0.0358& 51&        & -0.0182& 102& -0.0001\\
22&         & -0.0353& 52&        & -0.0176& 104& -0.0001\\
23&  -0.0348& -0.0347& 53&        & -0.0170& \textbf{106}& -3$\times 10^{-5}$\\
24&  -0.0342& -0.0342& 54&        & -0.0164& \textbf{108}& -8$\times 10^{-6}$\\
25&  -0.0337& -0.0336& 55&        & -0.0159& \textbf{110}& -9$\times 10^{-7}$\\
26&  -0.0331& -0.0330& 56&        & -0.0153& \textbf{111}& -8$\times 10^{-8}$\\
27&  -0.0325& -0.0325& 57&        & -0.0147& &        \\
28&  -0.0320& -0.0319& 58&        & -0.0142& &        \\
29&  -0.0314& -0.0313& 59&        & -0.0136& &        \\
\hline\hline
\end{tabular}}
\label{tab:2}
\end{table}

%%%%%%%%%%%%%%%%%%%%%%%%%%%%%%%%%%%%%%%%%%%%%

\section{CONCLUSIONS}

{   Our main result is a prediction of the rovibrational spectra, both energies and eigenfunctions in the Born-Oppenheimer approximation, which is done for the first time. The Lagrange mesh method which is used to solve the one-dimensional (nuclear) Schr\"odinger equation allows to obtain accurately the wave function of each rovibrational state. As a result, it is possible to calculate both the expectation values and matrix elements, hence, the transition probabilities.}

The potential energy curve for the chromium dimer Cr$_2$ is presented in the form
of a two-point P\'ade approximant which reproduces correctly the asymptotic behavior
at small ($R\rightarrow 0$)~\re{small_distances} and large ($R\rightarrow\infty$)~\re{large_distances} internuclear distances, it is
valid over the entire $R$ domain $R\in[0,\infty)$. The potential energy curve
reproduces 3-4 decimal digits when compared with the turning points
presented in~\cite{HL:2022}. This allows us to recover the position and the depth
of the minimum of the potential energy curve. {   It is evident that the knowledge
of the potential energy curve in the entire domain of internuclear distances $R \geq 0$
allows us to construct the eigenfunctions of low-lying states and calculate transition
amplitudes, in particular, the electric dipole transition probability $(0,0) \rar (0,1)$.
This will be done elsewhere.}

By solving the nuclear Schr\"odinger equation~\re{Hamiltonian} for the
most abundant chromium isotope $^{52}$Cr, the obtained vibrational spectrum
$E_{(\nu,0)}$ ($L=0$) of Cr$_2$ shows agreement in 3-4 decimal digits
with the experimental data on vibrational energies ~\cite{HL:2022}. In general,
the potential energy curve support 112 vibrational states $E_{(\nu,0)}$ ($\nu_{max}=111$)
and 313 rotational states $E_{(0,L)}$ ($L_{max}=312$) at $\nu=0$.
When comparison is possible, the absolute error in the vibrational spectrum
$E_{(\nu,0)}$ is $\lesssim 4 \times 10^{-4}$ {\it hartree} for all available experimental data.
In total, there are 19912 rovibrational states, out
of which 218 are weakly bound (close to the dissociation limit) with
energies $<10^{-4}$ {\it hartree}.

It should be noted that the presented approach allows us to explore the
isotopologues of the Cr$_2$ dimer by simply changing the reduced mass $\mu$
in the Schr\"odiner equation of the nuclear motion~\re{Hamiltonian}. Also
by changing the first coefficients accordingly in the expansions
\re{small_distances} and \re{large_distances} and by adding experimental or
{\it  ab initio} data the electronic potential curves for the first excited electronic
states of Cr$_2$ can be constructed. Needless to say that by taking the expansion
for large distances to
\begin{equation}
    \tilde{E}\ =\ -\frac{C_4}{R^4}\ +\ O\left(\frac{1}{R^6}\right)\ ,
\label{large_distances-ions}
\end{equation}
cf.\re{large_distances} one can study the positive/negative molecular ions
Cr$_2^{\pm}$.

In conclusion, we have to state that alternative experimental studies
to the Casey-Leopold one \cite{CL:1993} are needed to confirm and extend
their results.

%-------------------------------------------------------------------------------------

\section*{ACKNOWLEDGMENTS}
The authors thank T.~Rocha (UNAM, Mexico) and P.~Zuchowski (Torun, Poland) for bringing this problem
to our attention. Discussions with J.~Karwowski (Torun, Poland) are highly appreciated.
D.A.D. is grateful to CONACyT (Mexico) for a graduate scholarship.
The research is partially supported by CONACyT grant A1-S-17364 and DGAPA grant
IN113022 (Mexico).

%\newpage

%-------------------------------------------------------------------------------------

\end{document}